\documentclass[letterpaper, journal, twoside]{IEEEtran}

\usepackage{amsmath,amssymb,amsfonts}
\usepackage{algorithm}
\usepackage{algpseudocode}
\usepackage[caption=false,font=normalsize,labelfont=sf,textfont=sf]{subfig}
\usepackage{bm}
\usepackage{textcomp}
\usepackage{stfloats}
\usepackage{url}
\usepackage{verbatim}
\usepackage{graphicx}
\usepackage{siunitx}
\usepackage{balance}

\usepackage{amsthm}

\newtheorem{definition}{Definition}[section]

\newtheorem{assumption}{Assumption}

\usepackage{lipsum}
\usepackage{mathtools}

\usepackage[pdftex, plainpages=false,hypertexnames=true,pdfnewwindow=true,colorlinks=true,citecolor=blue,linkcolor=red,urlcolor=blue,filecolor=blue]{hyperref}%

\begin{document}

\title{\LARGE \bf Multi-UAVs end-to-end Distributed Trajectory Generation over Point Cloud Data  }

\author{Antonio Marino, Claudio Pacchierotti, Paolo Robuffo Giordano
\thanks{Manuscript received: February 26, 2024; Revised:
April 23, 2024; Accepted: June 15, 2024.}
\thanks{This paper was recommended for publication by Giuseppe Loianno upon evaluation of the Associate Editor and Reviewers’ comments.}
\thanks{A. Marino is with Univ Rennes, CNRS, Inria, IRISA -- Rennes, France. E-mail: antonio.marino@irisa.fr}
\thanks{C. Pacchierotti and P. Robuffo Giordano are with CNRS, Univ Rennes, Inria, IRISA -- Rennes, France. E-mail: \{claudio.pacchierotti,prg\}@irisa.fr}
\thanks{This work was supported by the ANR-20-CHIA-0017 project ``MULTISHARED''}
\thanks{Digital Object Identifier (DOI): see top of this page.}
}

\maketitle

\begin{abstract}
This paper introduces an end-to-end trajectory planning algorithm tailored for multi-UAV systems that generates collision-free trajectories in environments populated with both static and dynamic obstacles, leveraging point cloud data. Our approach consists of a 2-branch neural network fed with sensing and localization data, able to communicate intermediate learned features among the agents. One network branch crafts an initial collision-free trajectory estimate, while the other devises a neural collision constraint for subsequent optimization, ensuring trajectory continuity and adherence to physical actuation limits. Extensive simulations in challenging cluttered environments, involving up to 25 robots and 25\% obstacle density, show a collision avoidance success rate in the range of $100-85\%$. Finally, we introduce a saliency map computation method acting on the point cloud data, offering qualitative insights into our methodology.
\end{abstract}


\section{INTRODUCTION}
The domain of multi-agent Unmanned Aerial Vehicle (UAV) trajectory planning has garnered significant attention, given its diverse range of applications \cite{gu2018multiple, peng2022obstacle, gao2022meeting}. In these settings, planning algorithms play a pivotal role in calculating trajectories that are both safe and directed towards a defined goal. These algorithms have to consider the dynamic state of the environment and the presence of neighbouring agents. In practical applications involving drones, trajectory planning is crucial to navigate around obstacles and accommodate limitations in the drone's actuators~\cite{alcantara2021optimal,kim2016realization,zhao2021multi}.

Despite possessing more theoretical guarantees, centralized approaches are often less appealing than decentralized counterparts due to their computationally intensive nature and the dependence on full-state information of each robot at every algorithm iteration which renders them impractical for real-world execution~\cite{chen2021decentralized}. In contrast, decentralized planning not only demonstrates enhanced scalability but also provides robustness against potential failures in a centralized architecture~\cite{I-Magnus2017ControlMRSsurvey}.

For more effective coordinated planning, the planning algorithm must consider not only local sensing data but also the planning decisions of a few neighbouring team members~\cite{cortes2017coordinated}. Thus, communication emerges as a critical element in realizing distributed solutions for multi-agent systems. Within this context, one of the focuses of this article is to integrate local sensing and communication strategies to achieve end-to-end distributed multi-UAV motion planning.

For UAV navigation, learning-based control from point clouds or images offers computational advantages~\cite{loquercio2021learning, palossi201964, miera2023lidar}. Recently, distributed learning methods for multi-agent scenarios have emerged, using reinforcement learning~\cite{el2023reinforcement, liu2019reinforcement, batra2022decentralized} and Graph Neural Networks (GNNs)~\cite{li2021message, li2020graph}. GNNs, by exploiting the communication graph, excel in encoding distributed policies, as demonstrated by Blumenkamp et al.~\cite{blumenkamp2022framework}, who showed multi-robot coordination in narrow passages. However, learning-based control needs additional safety guarantees, as seen in GLAS~\cite{riviere2020glas}, which uses a local safety function to ensure stability and safety. Most results apply to static environments without considering real robot dynamics. A recent work~\cite{zhang2023neural} combines safety control strategies with learning over a graph of agents and obstacles, using a GNN-based network to predict control estimates and a constraint function for Control Barrier Function (CBF) optimization from LiDAR data.

However, all these methods are susceptible to trapping robots in local minima as they compute only a local control action. To explicitly tackle this problem, a contribution of this work is to propose a trajectory planning algorithm that takes into account spatio-temporal predictions and can avoid trapping the robots in deadlocks. Planning for multi-drone coordination inherently poses a high-dimensional challenge, even when safety is the sole requirement. Moreover, planning algorithms often necessitate access to map data for obstacle information, as noted in prior work~\cite{sabetghadam2022distributed}. This map should be updated online to tackle environmental changes which in large-scale environments can become cumbersome, bringing a computational bottleneck. Furthermore, optimization-based approaches may need to relax collision constraints as team density increases. Potentially, this relaxation leads to collisions, as discussed in~\cite{luis2020online} which reports real-time motion planning for a swarm of up to 20 drones. The recent literature~\cite{tordesillas2021mader, kondo2023robust, wang2022robust, park2022online} has presented notable results for robust multi-drone large-space travelling that take into account physical size, actuator limitations, alongside tracking disturbances, communication delay, and asynchronous communication. However, these algorithms typically rely on perfect sensing and tracking of obstacles at all times or a pre-available obstacle map, which, generally, is unavailable in real-world scenarios. Furthermore, these algorithms require tuning the map grid size to find collision-free paths and, despite implementing a decentralized approach, each drone must communicate with all other drones in the team, which reduces scalability.

In this paper, we present a novel data-driven approach for distributed trajectory generation and safety collision avoidance under LiDAR-based observations. Our main contributions are:
\begin{itemize}
    \item A decentralized and asynchronous end-to-end trajectory planning method using attention-based GNNs to learn a decentralized policy from point cloud data. Unlike existing methods requiring prior knowledge of the environment, our approach allows each drone to plan its trajectory directly from sensed data.
    \item An optimization process that generates collision-free trajectories by predicting collision constraints from point cloud data and inter-drone communication. This method allows real-time updates and adjustments, ensuring higher reliability and safety in dynamic, uncertain environments.
    \item Leveraging the point cloud saliency map computed by a variation of VisBackProp~\cite{bojarski2016visualbackprop}, we propose a qualitative analysis of the predicted trajectories and how the point cloud contributes to it.
\end{itemize}


\section{PROBLEM STATEMENT}
\label{sec:prob-statement}
Consider a distributed trajectory generation problem for a set of UAVs $ V \coloneq \{1, \dots N \}$   modelled at time $t$ in $SE(3)$ with second-order dynamics. The UAV team operates within a three-dimensional workspace containing static and dynamic obstacles $\mathcal{O}_t$. Each drone's objective is to navigate the environment toward its target without colliding with other agents. A team mission is the collective traversing of the space such that every drone satisfies its objective. We make the following assumptions 
\begin{assumption}
Each drone is controlled via a trajectory-tracking algorithm and equipped with a 3D range sensor, such as a LiDAR, with $360^{\circ}$ field of view.
\end{assumption}
\begin{assumption}
At time $t$, we assume perfect state estimation, i.e. the quaternion orientation $\bm{quat}_i(t)$ and the triplets of position $\bm{r}_i(t)$, velocity $\bm{v}_i(t)$ in world frame are available to agent~$i$.
\end{assumption}
\begin{assumption}
Each drone can communicate with a limited number of neighbours at a given frequency without communication loss  
\end{assumption}

The observation data for a drone $i$ is denoted by $pc_i \in R^{m \times 3}$, which includes the relative positions of the $m$ ``hit'' points in the scanned environment. The trajectory generated must comply with the following constraints: 
\begin{itemize}
    \item \textbf{collision avoidance}: each pair of drones and drone-obstacle maintains a safety distance of $2d$ throughout the whole trajectory, where $d > 0$ is the radius of a sphere containing the physical body of the agents. 
    \item \textbf{limited sensing and communication}: each agent has a limited sensing and communication range radius $R$. We define the neighbours of agent $i$ as $\mathcal{N}_i = \{ j \in V~|~ \| \bm{r}_i - \bm{r}_j \|_2 \leq R, j \neq i \} $; therefore, the agents can only sense other agents or obstacles inside a sphere of radius $R$ originating on the agent.
    \item  \textbf{physical limitations}: The generated trajectory must satisfy the drone's velocity and acceleration limits and must be smooth and continuous relative to the actual state.     
\end{itemize}

\section{METHOD}
\label{sec:method}
\begin{figure*}[t]
    \centering    
    \includegraphics[width=0.9\textwidth]{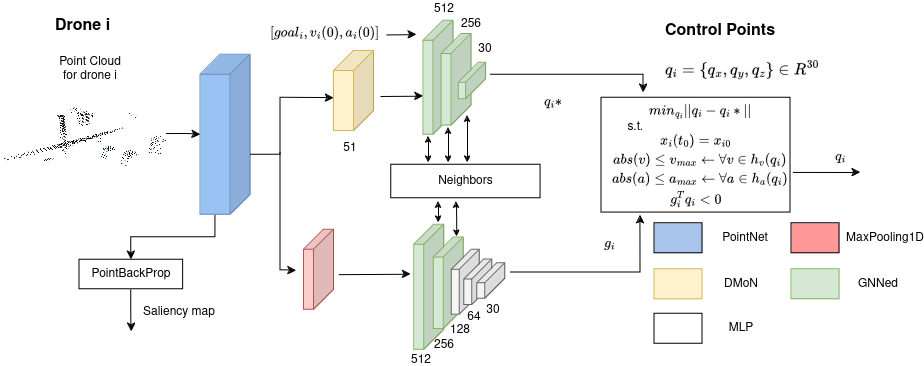}
    \caption{neural network architecture deployed on each drone.}
    \label{fig:net_architecture}
\end{figure*}

This section presents the proposed policy as a neural network that fulfils the above requirements. We start by presenting the trajectory encoding used by the proposed approach. Then, we present the privileged expert used for training and how we process the input features. Subsequently, we describe the neural network architecture depicted in Fig.~\ref{fig:net_architecture} and deployed on the drone composed of input features processing, communicated variables aggregation and the optimization layer to cope with physical limitations. Finally, we describe the algorithm to compute the saliency map, used to analyze the neural network behaviour in the results.
\subsection{Trajectory Representation}
The trajectory used by each drone is a polynomial-based curve, e.g., B-Splines, which allows us to impose velocity and higher derivatives limitation through linear constraints on the trajectory Control Points (CP). In fact, the convex hull of the curve CP leads to the representation of the outer trajectory polyhedron and, by imposing constraints on the CP, we ensure they are satisfied throughout the trajectory. 

In this paper, we adopted a MINVO basis~\cite{tordesillas2022minvo} expression for the curve. MINVO bases are recently introduced in the literature and, compared to B-spline or Bernstein bases, form a simplex $\mathcal{M}$ enclosing the given polynomial curve with minimum volume by construction. Hence, these bases lead to less conservative polyhedron representations.
Given a polynomial order, we can pass from MINVO to B-Spline CP and vice-versa by a linear transformation. For these reasons, it is ideal to generate and apply constraints on MINVO CP for trajectory descriptions. In particular, we define two linear mappings, $h_v(\cdot)$ and $h_a(\cdot)$, to pass from MINVO CPs position to their velocity and acceleration, respectively.      

\subsection{Privileged Expert}
Our trajectory planner is trained via privileged learning~\cite{chen2020learning}. Specifically, we generate a dataset from an expert controller: MADER~\cite{tordesillas2021mader}. MADER employs a decentralized and asynchronous approach to generate feasible and safe trajectories, leveraging a combination of path planning and Quadratic Programming (QP) optimization. We can consider this algorithm as a privileged expert for two reasons: first, MADER exploits comprehensive knowledge of the physical dimensions of obstacles and drones, as well as the trajectories of both, which is a piece of information typically unavailable in real-world scenarios; second, MADER ensures to find feasible trajectories only for unlimited time budget available in simulation. The optimization proposed by MADER focuses on determining the separation planes between the drone and potential obstacles throughout the trajectory, facilitating motion within a safe region. Moreover, the same reasoning is applied to trajectories committed by the other drones, leveraging the trajectory outer simplex as a cluttered area. However, the growth of the decision variables is proportional to the number of obstacles and other drones within the decision horizon. This renders the optimization more difficult to solve within a limited time budget, leading to potential infeasibility and hazardous halts. However, in simulation where full obstacle information is available and no time constraints are present, we can fully exploit MADER to generate a dataset of safe trajectories.

In our approach, to construct a dataset, we chose MADER over centralized alternatives as its features of asynchrony and decentralization align well with the objectives of our trajectory planner. Moreover, learning over an already decentralised expert policy eases the training process. After the training process, at execution time, the learned policy does not rely on any privileged information and synthesizes the trajectory from sensor inputs only.

\subsection{Neural Network Input Features}
Every drone $i$ in the team shares the same neural network architecture that accepts as input raw point clouds normalized in a unitary sphere and transformed into the world frame using $\bm{quat}_i$.  Additionally, the proposed neural network uses also the current velocity, $\bm{v}_i$, and the current desired acceleration, $\bm{a}_i$, normalized by the maximum velocity $\bm{v}_{max}$ and the maximum acceleration $\bm{a}_{max}$ allowed.  We also include the goal location ($\bm{goal}_i$) in the drone $i$ frame, projected on the sensing sphere with radius $R$ if the distance to the goal is greater than $R$. Then, the $\bm{goal}_i$ is divided by $R$ to have coordinates in the range $[-1,1]$. Drone $i$ can communicate intermediate neural network features with drones in its neighbourhood $\mathcal{N}_i$. We enclose the drone initial conditions in $\bm{x}_{i0} = [ \bm{0}_{3}, \bm{v}_i, \bm{a}_i]$ for a generated trajectory starting from the ego location, $\bm{0}_{3}=[0,0,0]$. In the final deployment, we add to the learnt control point $\bm{q}$ the current drone location $\bm{r}_i(t)$ to track the trajectory.
\subsection{Neural Network Structure}
The neural network consists of two main branches that process the point cloud $\bm{pc}_i$, combine it with localization data $[\bm{goal}_i, \bm{v}_i, \bm{a}_i]$, and communicate features to compute a trajectory guess vector $\bm{q}_i^{*}$ (described by MINVO CPs) and a vector of collision coefficients $\bm{g}_i$ for drone $i$. The latter is used to define a linear combination of CPs composing $\bm{q}_i$, thereby classifying its safety. Specifically,

\begin{definition}
\label{def:safe_set}
    Given $pc \coloneq [pc_1^T \dots pc_N^T]^T$, the joined point clouds from $N$ drones, for drone $i$, the scalar value $\bm{g}_i^T\bm{q}_i$ is $\bm{g}_i^T\bm{q}_i < 0$ if and only if the control points vector $\bm{q}_i$  belongs to the safe set $\mathcal{S}_i$, where $\mathcal{S}_i = \{ \bm{q}_i ~|~\|\bm{p}_i - \bm{p}_j\|_2>2d~, \bm{p}_i \in \mathcal{M}(\bm{q}_i), \bm{p}_j \in pc \}$ and $\mathcal{M}(\bm{q}_i)$ is the outer simplex of $\bm{q}_i$.   
\end{definition}
The term $\bm{g}_i^T\bm{q}_i$ can be used as a linear constraint to generate a safe trajectory. The collision coefficients $\bm{g}_i$ serve to fortify the learned policies against collisions but also to ensure their applicability and reliability in unseen scenarios where the expert policy is unavailable or not included in the dataset. We formulate a QP layer as the final stage within the neural network that optimizes $\bm{q}_i$ to closely align with the initial guess $\bm{q}_i^*$, while concurrently ensuring adherence to predefined maximum drone velocity ($\bm{v}_{max}$) and acceleration ($\bm{a}_{max}$) constraints. This combination not only refines trajectory predictions but also strengthens the system's resilience in navigating complex environments. The QP formulation is the following:
\begin{equation}
    {\small
    \begin{aligned}
        &\min\limits_{\bm{q}_i} \| \bm{q}_i - \bm{q}_i^{*} \|_2^2   \\
        \text{s.t.} & \\
            & \bm{x}_i(t_0) = \bm{x}_{i0} \\
            & abs(\bm{v}) \leq v_{max} \xleftarrow{} \forall \bm{v} \in h_v(\bm{q}_i) \\
            & abs(\bm{a}) \leq a_{max} \xleftarrow{} \forall \bm{a} \in h_a(\bm{q}_i) \\
            & \bm{g}_i^T\bm{q}_i \leq 0 \\  
    \end{aligned}}%
    \label{eq:QP}
\end{equation}
The point cloud is processed by a PointNet layer~\cite{charles2017pointnet}  comprising three filters of $64, 128, 256$, respectively. PointNet employs sequences of point transformations through a compact transformation network and 1D-CNN with a unitary kernel size to generate spatially-permutation invariant features. The output of this layer, $m\times256$ with $m$ points, splits to serve distinct purposes within the collision coefficients branch and the trajectory generation branch. To generate $\bm{g}_i$, i.e., assigning a safety score to trajectories crossing the environment, we require global features extracted from the point cloud. Consequently, the PointNet output is transformed into a $256$-dimensional vector through global max pooling over the features.\\ 
Conversely, within the trajectory generation branch, our approach involves further clustering of the point cloud based on the features learned by PointNet. Clusters prove advantageous in classifying spatial regions that hold crucial information for trajectory generation. Clustering was already adopted in the design of PointNet++, which consists of spatial clusterization coupled with smaller PointNet modules. We adopted DMoN~\cite{tsitsulin2023graph}, a clustering methodology based on a graph neural network, approximating spectral modularity maximization to recover high-quality features and spatial-based clusters. Unlike PointNet++, DMoN's approach does not solely rely on spatial displacement and does not require a nested architecture of PointNets. The adjacency matrix required by DMoN is represented as a binary sparse map encoding the spatial proximity of the points. This strategic combination of PointNet and DMoN very well captures intricate spatial structures for trajectory generation tasks, as we show in the results of Sec.~\ref{sec:experiments}. We chose a number of clusters ($51$) that, together with the localization data, form a $64$-dimensional vector for the next layer. \\
Both branches exchange the local features with the neighbouring agents using a message-aware graph attention network (MAGAT)~\cite{li2021message}. Assuming that each drone $i$ can communicate with its set of neighbours $\mathcal{N}_i$, the communication graph can be represented by a binary sparse matrix $\bm{A}$. We let $\bm{A} \in \mathbb{R}^{N \times N}$ be the adjacency matrix of the communication graph. Given a graph signal $\bm{x} \in \mathbb{R}^{N \times F}$ distributed over the drones, we can define a graph filter as
\begin{equation}
   H_{\bm{A}}(\bm{x}) = \sum_{k=0}^K \bm{A}^k \bm{x} \bm{H}_k.
    \label{eq:filter}
\end{equation}
which combines the elements of $\bm{x}$ over the adjacency matrix of the communication graph and applies the graph filter weights $\bm{H}_k \in \mathbb{R}^{F \times F'}$. The quantity $K\geq1$ represents the filter length, which implies repeated $1$-hop communications over the graph. Therefore, the filter can be executed distributively over the graph. The filter in eq.~\eqref{eq:filter} transforms the graph signal from an $F$-features space into a signal of an $F'$-features space. MAGAT enhances this framework by incorporating an attention mechanism to weigh the relative importance of drone features. Specifically:
\begin{equation}
{\small
 \begin{aligned}
    H_{\bm{A\mathcal{E}}}(\bm{x}) &= \sum_{k=0}^K (\bm{\mathcal{E}} \circ \bm{A})^k \bm{x} \bm{H}_k. \\
    \mathcal{E}_{ij} &= \frac{exp(LeakyReLU(\bm{x}_i^{T}W\bm{x}_j))}{\sum_{k \in \mathcal{N}_i}exp(LeakyReLU(\bm{x}_i^{T}W\bm{x}_k))}
    \label{eq:MAGAT-filter}
    \end{aligned}
    }%
\end{equation}
where $\bm{\mathcal{E}}$ is the matrix of attention weights. Considering the need to communicate graph signals over the network, the size of $F$ affects network congestion. To address this, we extend the scheme by introducing an encoding-decoding mechanism to compress the graph signal before communication and reconstruct its original dimension afterwards. This involves point-wise learnable encoding and decoding functions, denoted as ${ e_{\theta}(\cdot):\mathbb{R}^{F} \rightarrow \mathbb{R}^{G}}$ and ${d_{\theta}(\cdot):\mathbb{R}^{G} \rightarrow \mathbb{R}^{F}}$, respectively: 
\begin{equation}
 \begin{aligned}
    H_{\bm{Aed}}(\bm{x}) &= \sum_{k=0}^K d_{\theta}((\bm{\mathcal{E}} \circ \bm{A})^k e_{\theta}(\bm{x})) \bm{H}_k. 
    \label{eq:MAGAT-filter-ed}
    \end{aligned}
\end{equation}
Here, $G << F$ reduces the number of communicated variables, with $e_{\theta}$ and $d_{\theta}$ functions implemented as MLP with ReLU activation functions. The resulting GNNed layer,
\begin{equation}
 \begin{aligned}
    \bm{x} = ReLU(H_{\bm{Aed}}(\bm{x}))
    \label{eq:GNN-ed}
    \end{aligned}
\end{equation}
is employed $3$ times in the trajectory generation branch (with filter dimensions $512, 256, 30$) and twice in the collision branch (with filter dimensions $512, 256$), totalling $L=5$ layers. Set parameters include $K=1$ and $G=5$. We can use the unit-delay communication model~\cite{gama2022synthesizing} and communicate unit-time delayed signals to compute the graph filters in~\eqref{eq:MAGAT-filter-ed} in one shot, by sending~${[\bm{x}_i(t), \hdots, A_i^{K-1}\bm{x}(t-K-1)]}$ for each agent. This model allows releasing a new output at each communication iteration but at the cost of more communicated variables. Therefore, each drone communicates $LGK$ variables, i.e. $25$ in our implementation.

\subsection{Point Cloud Saliency Map}
A saliency or attention map, denoted as $s = [0,1]^m$, serves as a feature map unveiling the relevance of input data in the decision-making process. This map is instrumental in inspecting and interpreting neural networks, particularly under distribution shifts. Existing studies on point cloud saliency maps~\cite{jiang2023c2spoint, zheng2019pointcloud} predominantly rely on gradient-based methods or incorporate additional neural network modules. As highlighted by the authors of VisBackProp~\cite{bojarski2016visualbackprop}, for autonomous navigation, it is convenient to dispose of a gradient-free method that can be evaluated online.  Addressing this gap, we applied the same reasoning of VisBackProp to the PointNet layer in our architecture, proposing a variation of VisBackProp that can handle point cloud. We focus on the PointNet layer in our architecture, as it processes the point cloud and is dynamically shaped during training to enhance the representation of points for trajectory generation. The 1D-CNNs composing the PointNet draw a direct parallelism with VisBackProp that is applied to 2D-CNNs. For a PointNet made of $P$ sequences of feature transformations and 1D-CNN, we save the average map $\bar{h}_i$ computed over the features generated after each sequence. Starting from $\Bar{h}_P$, we loop over the list of $\bar{h}_i$ by multiplying the actual element with previous element of the list and normalize between $[0,1]$ after each multiplication. The algorithm is summarized in Algorithm~\ref{alg:pointbackprop}.

\begin{algorithm} [t]
\caption{ PointBackProp}\label{alg:pointbackprop}
\begin{algorithmic}
\Require point cloud $(pt \in \mathbb{R}^{m \times 3})$, PointNet
\State $h_1, h_2, \dots, h_P \gets$ get the $P$ intermediate features computed by PointNet 
\State $\bar{h}_1,\bar{h}_2, \dots, \bar{h}_P \gets$ compute average of the features
\State $s := \bar{h}_P$ \Comment{ saliency map $s = [0,1]^m$}
\For {i = P-1 ... 1}
    \State $s:= \bar{h}_i \cdot$ $s$
    \State Normalize s between $\{0, 1 \}$ 
\EndFor
\State compose point cloud values $\{s,pt\}$
\end{algorithmic}
\end{algorithm} 
\section{TRAINING}
\label{sec:training}
\begin{algorithm} [t]
\caption{ Training  Algorithm}\label{alg:training}
\begin{algorithmic}
\Require $ \mathcal{D}$ batch size $b$
\For {point cloud in $ \mathcal{D} $}
    \State Combine point clouds of the $N$ drones
    \State Generate random trajectories $\hat{q}$
    \State Solve QP to constraint $\hat{q}$ with the initial conditions and physical limitations
    \State  add $\hat{q}$ to $\mathcal{S}_i$ according to the Definition~\ref{def:safe_set}
    \State Append trajectories and collision indexes to the dataset $\hat{\mathcal{D}} \gets \hat{\mathcal{D}} \cup \{\hat{q}\}$
\EndFor
\State calculate adjacency matrix of drones $A$ 
\Comment{This step is only needed in centralized training}

\State calculate adjacency of point clouds $A_{pt}$
\For {$i= 1 \dots$ epochs}
    \State collect $b$ batches $\{ pt, v, a, quat, goal, q \} \gets \mathcal{D}$
    \State compute $q*,g \gets$ model 
    \If{$i > epochs_i$}
        \State prepare QP in eq \eqref{eq:QP} with constraint $g^Tq* < 0$  
        \State Solve QP $\xrightarrow{} q $
        \If{QP feasible}
            \State $q \xrightarrow{} q* $
        \EndIf
    \EndIf
    \State extract $\{\hat{q}\}$ in $\hat{\mathcal{D}}$ 
    \State Compute constraint $L_g(g,\hat{q})$ using eq~\eqref{eq:loss_collision}
    \State Compute loss $L_q(q,q*)$ in eq~\eqref{eq:loss_traj}
    \State Update model weights
\EndFor
\end{algorithmic}
\end{algorithm}
\begin{figure}[t]
    \centering
    \includegraphics[scale=0.15]{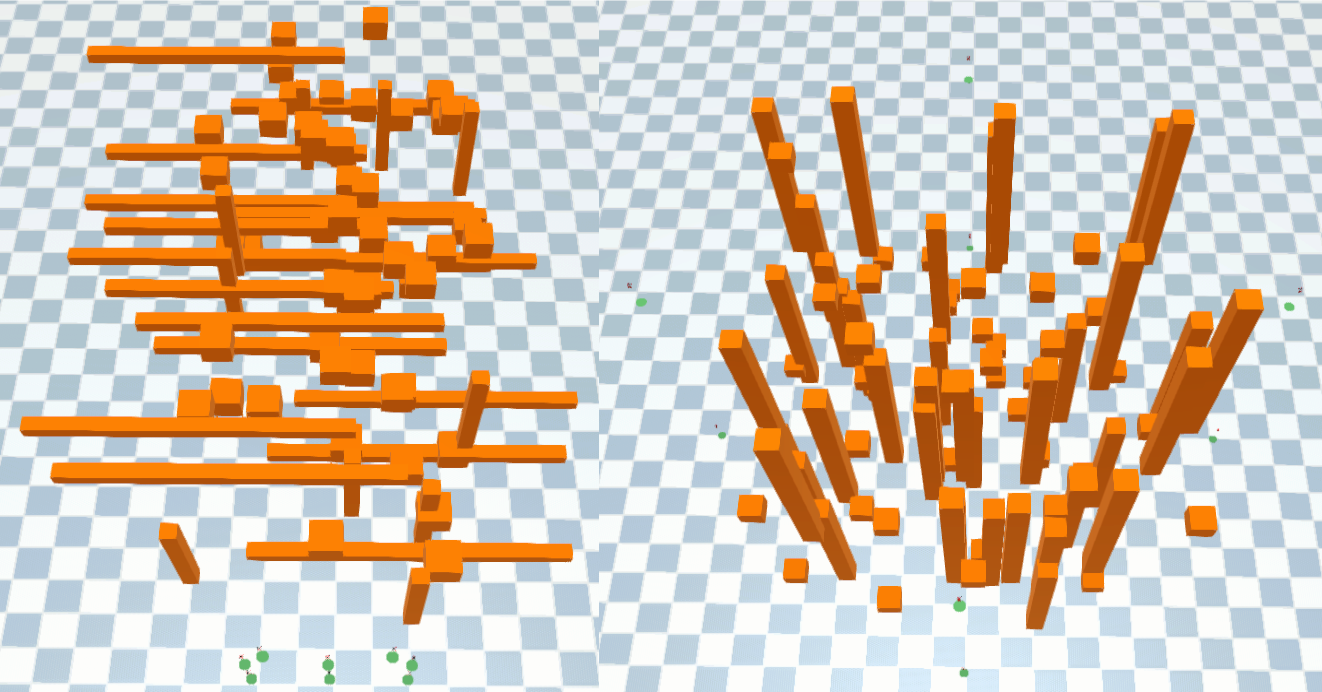}
    \caption{Experimental environments: dynamic corridor (right) and dynamic forest (left).   }
    \label{fig:test_scenarios}
\end{figure}
Our model training dataset $\mathcal{D}$ encompasses UAV around 16k trajectories under the control of the expert controller within a simulated environment. The simulation environment incorporates a $10\%$ obstacle density organized in a forest-like setting and confined within a sphere of $10$\unit{m} radius centred in the world origin. The obstacles are partitioned equally into static and dynamic elements, with the dynamic obstacles following a three-dimensional trefoil knot motion spanning a width of $1$\unit{m}. For each mission, we randomly placed $8$ drones around the surface of the $10$-\unit{m} sphere, with each drone target located on the opposite side of the sphere surface, forcing the traversal of the sphere centre. The sensing/communication radius is fixed at $R=4$~\unit{m}. At a sampling rate of $15$Hz, we capture point cloud data, localization information, and current control trajectory in MINVO control points, starting from the current drone location. The trajectories are saved as clamped spline with $10$ equally spaced internal knots for each axis starting from the current timestamp $t_0$ to $t_f$. Therefore, the trajectories are defined solely by $10$ CPs for each axis for segment polynomials of degree $3$. This helps the training process by focusing on learning the control points. In addition, we generate 560k random trajectories $\hat{q}$ offline, originating from the initial robot location, and categorize them as successful or unsuccessful based on definition~\ref{def:safe_set} within the joint point cloud space, i.e. successful if the trajectory does not collide with obstacles or agent. Moreover, we introduce a diminishing safety distance $d$ along the trajectory, ranging from the drone's actual size at the trajectory's initiation to $0$ at its termination. This approach optimizes the safe set by progressively reducing conservatism, thereby prioritizing the safety of the trajectory's initial location:
\begin{equation}
        L_q =  \sum_{i \in V} \frac{1}{30}||q_i -q_i^*||^2_2 \quad q_i \in \mathcal{D},
\label{eq:loss_traj}
\end{equation}

\begin{equation}
        L_g =  \sum_{i \in V}\sum_{\hat{q}_i \in \mathcal{S}_i } [\bm{g}_i^{T}\hat{\bm{q}}_i]^{+} \sum_{\hat{q}_i \notin \mathcal{S}_i} [-\bm{g}_i^{T}\hat{\bm{q}}_i]^{+},
\label{eq:loss_collision}
\end{equation}
where $[\cdot]^{+} = Softplus(\cdot)$ stands for a continuous ReLU function ensuring strict constraint satisfaction. We split the dataset into training, validation, and testing at ratios of 0.6, 0.3,
and 0.1 respectively. We solve the imitation learning problem by using behavioural cloning with ADAM optimizer~\cite{kingma2014adam} employing a learning rate $1e-3$ and forgetting factors $0.9$ and $0.999$. We trained for $200$ epochs of which in the first $50=epochs_i$ the loss of eq.~\eqref{eq:loss_traj} is computed on the trajectory guess, without solving the QP. This solution helps to ease the learning and predict a good initial guess. The rest of the $150$ epochs use the QP optimization as last layer as explain in Sec.~\ref{sec:method}. If the QP is unfeasible, we used the initial guess to train our neural network. We exploit OptNet~\cite{amos2017optnet} to realize a differentiable quadratic programming layer and use classic backpropagation methods during training. We summarize the training process in the Algorithm ~\ref{alg:training}.

\section{EXPERIMENTS}
\label{sec:experiments}
\begin{figure}[t]
     \centering
     \includegraphics[width=0.48\textwidth]{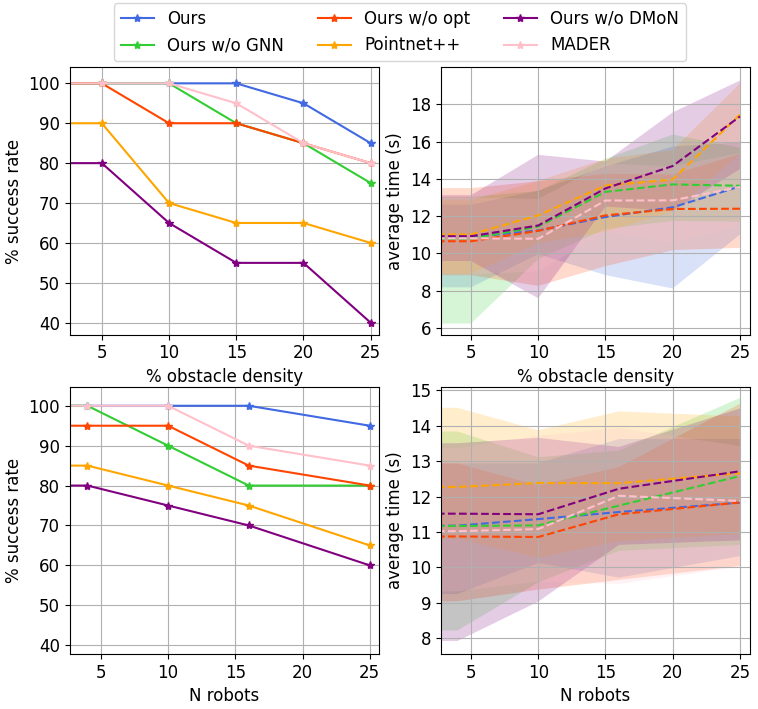}
\caption{ Success rate and safety rate increasing obstacles and agents in the range of obstacle density [$ 5\% - 25\% $] and number of robots between [$5-25$] for our approach, ours without GNN, ours without collision constraint, ours with Pointnet++, ours without DMoN and MADER algorithm. }
\label{fig:success_rate}
\end{figure}
\begin{figure}[t]    
     \centering
     \includegraphics[scale=0.19]{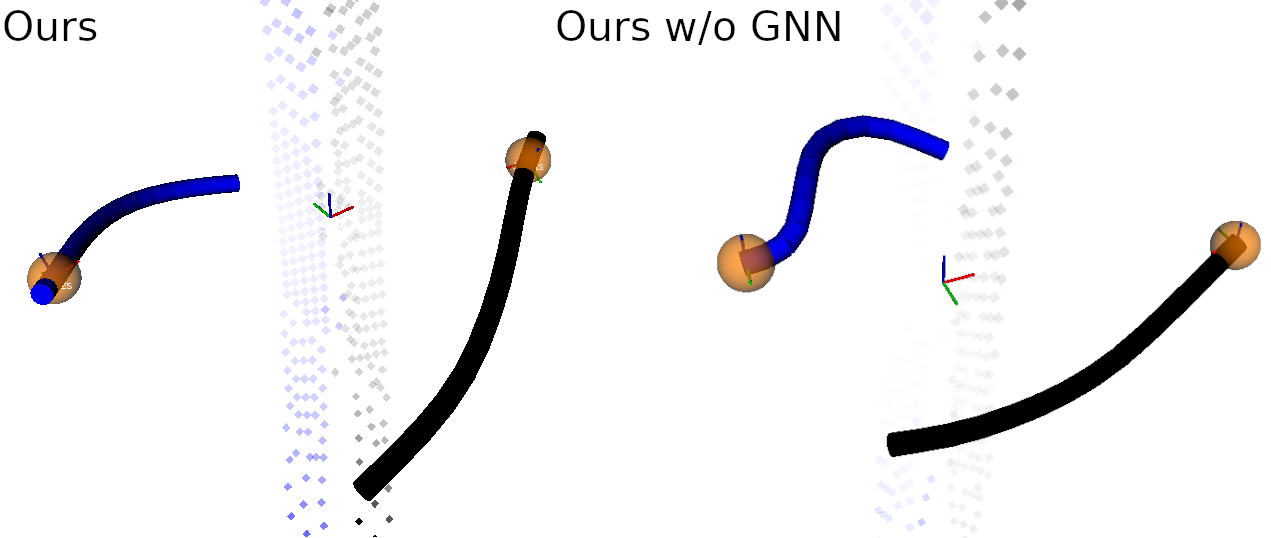}
\caption{Saliency map in a scenario with two drones for our approach and ours without GNN. The two drones (black and blue) sense each other and a pilar through their point cloud while moving toward the target. The saliency map is displaced by the alpha channel of the points.}
\label{fig:saliency2}
\end{figure}
\begin{figure}[t]
     \centering
     \includegraphics[width=0.48\textwidth]{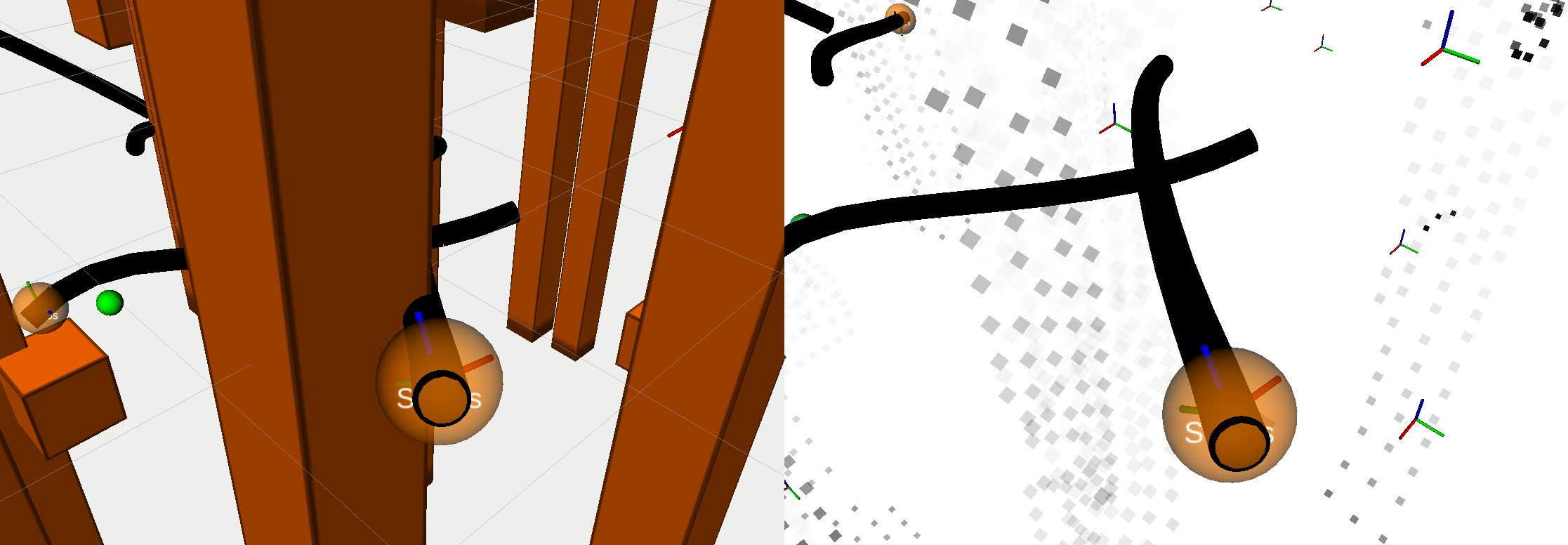}
\caption{Saliency map (on the left) of collision case using our approach for a trajectory starting from the drone and traversing the obstacle.}
\label{fig:uav-unsuccess}
\end{figure}

All simulations and training were conducted on a machine running Ubuntu 22.04. with Intel Core i7-9750H @ 2.60GHz CPU, Nvidia RTX 2080Ti and 32G RAM. The communication between the drones is employed through ROS with a communication rate of $100Hz$ while a new trajectory is calculated when a new point cloud is sensed at $15Hz$. As specified in the assumption, we do not consider communication loss but, being the trajectory computation at a lower rate, we can guarantee a communication delay margin of $0.056$~\unit{s}.
We first provide an ablation study and analysis results for a general quadrotor model with a perfect controller tracking trajectories with a maximum twist equal to $3.5$~\unit{m/s} in any directions and maximum acceleration equal to $[20,20,9.6]$~\unit{m/{s^2}}. We assume the quadrotor size to be confined in a sphere of radius $d=0.15\unit{m}$. Moreover, we chose $t_f=1$\unit{s} which offers a good compromise between knots resolution and prediction horizon. in this setting, the obstacle reaches a maximum velocity of $2$~\unit{m/s}. Then, we provide experimental results using Crazyflies drones in a physical simulation with maximum velocity $[1.0, 1.0, 1.0]$~\unit{m/s} and maximum acceleration $[2.0, 2.0, 2.0]$~\unit{m/s^2}. We leveraged a simulator built in Unity+Mujoco and software in the loop (SITL)\footnote[1]{https://gitlab.inria.fr/amarino/crazyswarm2\_unity\_sim} to mimic the real drone behaviour. In this setting, we reduce the obstacle maximum velocity to $1$~\unit{m/s} and $t_f=3$\unit{s} because of the drone's reduced velocity.

We consider two experimental conditions: a variable number of agents in a range of $[4,25]$, with dynamic and static obstacle density of $10\%$ of the space; a fixed number of agents to $8$ and increasing obstacle density from $5\%$ to $25\%$. In both conditions, we evaluate the average success rate and average travel time for the agents to reach their respective targets. We tested the approach in two settings shown in Fig.~\ref{fig:test_scenarios}: the forest-like used during training and a corridor-like environment $8\unit{m} \times 20\unit{m}$ including also horizontal pillars. In the second setting, the drones are required to traverse the corridor from end to end.
For each experimental condition (number of drones in team, obstacle density), we carried out $50$ repetitions of the travelling mission, equally distributed in the two scenarios. We consider a mission to be successful if all the agents reach their target without colliding. We proceeded with an ablation study to evaluate our approach: 
\begin{itemize}
    \item \textit{ours w/o GNN}: we replaced GNN layers with normal MLP, to evaluate the impact of communication over the predictions.
    \item \textit{ours w/o opt}: we did not use the predicted constraint $\bm{g}$ to test the impact of this introduced feature.
    \item \textit{Pointnet++}: we replaced our solution of Pointnet-DMoN with Pointnet++ and max pooling.
    \item \textit{ours w/o DMoN}: we removed DMoN clustering and used max pooling to predict the trajectory guess, as for the constraint prediction branch.
\end{itemize}
Additionally, for comparison, we also consider MADER, providing full obstacle trajectories but a limited time budget of $0.35$\unit{s} for each optimization according to the original paper to guarantee a real-time execution. Our approach reaches an average computational time of $1.3$\unit{ms} with a maximum of $1500$ points sensed. The computational time in real scenarios can change based on the LiDAR sensor resolution and the solver used for the last optimization layer which might be different from OptNet as, after the training, there is no need for the optimization to be differentiable. Note that, the sensing frequency dominates the computational time.

\subsection{Results}
\begin{figure}
    \centering
    \includegraphics[scale=0.48]{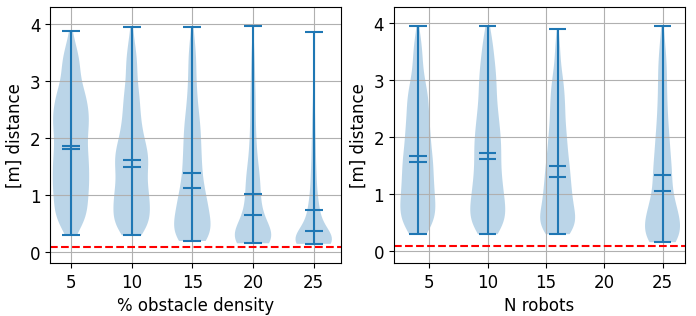}
    \caption{Distance from points sensed in a range of $4$~\unit{m} for flight tests with Crazyflie drones. The red dashed lines denote the physical safe distance.}
    \label{fig:crazyflie_flight_test}
\end{figure}

We present an analysis of the average success rate and average travel time on successful trajectories, as illustrated in Fig.~\ref{fig:success_rate}. Our proposed neural network consistently achieves a remarkable success rate of $100\%$ for obstacle densities up to $15\%$ and $16$ robots, gradually decreasing to $85\%$ for the highest obstacle density of $25\%$. Notably, the effectiveness of our model is closely tied to point cloud processing, as evidenced by the success rates of \textit{Pointnet++} and \textit{ours w/o DMoN}, ranging from $90\%$ to $40\%$. The clustering capabilities of \textit{Pointnet++} contribute to an average $8.3\%$ higher success rate compared to \textit{ours w/o DMoN}, emphasizing its impact on the environment processing.

When the model does not exploit the learnt collision constraint $g$, it encounters difficulties in finding collision-free trajectories, especially in high-constraint spaces populated by both cooperative and non-cooperative agents where the success rate diminishes to $80\%$ for an obstacle density of $25\%$ and $25$ robots. The integration of GNN enhances the model's resilience to collisions and traversal time as the number of drones increases, showcasing the advantages of cooperative trajectory predictions. Our approach shows advantages also compared to MADER with a high density of obstacles and robots primarily because when the optimization in MADER does not find a feasible trajectory the algorithm keeps using the previously computed trajectory. When the drone reaches the end of the trajectory, it breaks remaining exposed to obstacle collision until the optimization becomes feasible again because of the dynamical changes in the environment. Notably, \textit{ours w/o opt} and MADER reach the same success rate of $85\%$ with $25\%$ obstacle density which we speculate is due to the drone coordination through GNN. 

Our approach exhibits the lowest average travel time of $11.8$\unit{s} with $25$ robots even if MADER and \textit{ours w/o opt} find faster trajectories to traverse the space with fewer drones ($[5-10]$) and travel times comparable to \textit{ours} as obstacle density increases. Moreover \textit{ours w/o opt} has the lowest travel time when the obstacle increases with $12.4\unit{s} \pm 2.1$, at the expense of less safe trajectories. In contrast, \textit{Pointnet++} and \textit{ours w/o DMoN} have similar travel times, approximately $17.3\unit{s}$ with a high variance of $2\unit{s}$, facing drone deadlock situations or predicting longer paths to reach the goal. 

Our approach reaches a similar average time to MADER as we used it to generate the dataset. However, we note that MADER does not generate global optimal time trajectories due to its asynchronous communication strategy, aiding the first drone committing the trajectory. As a result, subsequent drones must adapt to avoid collisions, resulting in non-optimal trajectories. 


We use Algorithm~\ref{alg:pointbackprop} to generate saliency maps, which help interpret the network's behaviour and provide insights into the point cloud's contribution to trajectory prediction. The saliency map allows us to qualitatively assess how well the neural network uses the sensing data for decision-making. We expect that the relevance of each point in the cloud is influenced by its distance from the drone, as the network predicts spatio-temporal commands and potential collisions around the drone, independent of the goal location. Initially, we focus our analysis on a scenario involving two drones and a static obstacle, as illustrated in Fig.~\ref{fig:saliency2}. To highlight the contribution of different observations, we assign distinct colours (blue and black) to points sensed by individual drones, with the alpha channel representing saliency values—more transparent points indicate less impact on predictions. It is evident from the saliency map that each drone perceives the obstacle from only one side. As expected, points closer to the drones are more transparent since the first part of the trajectory is predefined by the continuity with the current motion. Compared to \textit{ours w/o GNN}, the saliency map for our approach shows a balanced distribution of points between the two drones, resulting in more uniform trajectories across the environment. In contrast, when deploying \textit{ours w/o GNN}, we note an unbalanced use of the drone point cloud with the black drone relying more heavily on its sensed points than the blue drone, leading to distinct behaviours. This discrepancy indicates more potential conflicts between the drones due to differences in the perceptions of the environment, while GNN helps to reach coherent observations across different viewpoints. Additionally, saliency maps serve as valuable tools for analyzing failure cases. Figure~\ref{fig:uav-unsuccess} showcases a collision scenario with \textit{ours}. The trajectory intersects a pillar, with its points appearing transparent in the saliency map, suggesting that the network is "blind" to the obstacle, although perceived through the point cloud. Conversely, other obstacles are clearly visible and incorporated into the trajectory planning.
\subsection{Physical Simulation}
In a physical simulator, We tested our approach with Crazyflie drones. We recorded a mission for each experimental scenario: obstacle density ranging from $5\%$ to $25\%$ and numbers of agents ranging from $4$ to $25$. Moreover, as for the non-physical simulation, the missions were conducted for the two scenarios of a dynamic forest and a dynamic corridor. Figure~\ref{fig:crazyflie_flight_test} illustrates the distribution of accumulated drone-to-obstacle distances during the flights. Notably, no collisions were observed, as all recorded distances remained above the safety threshold of 0.1 meters.

\section{CONCLUSION} 
\label{sec:conclusions}
This work introduces a decentralized end-to-end trajectory planner that addresses static obstacles, dynamic obstacles, and other agents. By learning a collision constraint alongside the trajectory, we ensure the safety and dynamic feasibility of the trajectories in a QP framework. We extensively validate our approach in simulation against variations of our approach and demonstrate robustness to team scalability and varying obstacle densities in the environment. Additionally, we derived an algorithm to compute the saliency map of the point cloud, which we used as a tool to interpret and understand both failure and success cases. Future work could explore how to exploit the saliency map to enhance prediction performance and consider alternative learning frameworks that do not rely on privileged experts. Furthermore, future research will examine additional robustness metrics under conditions such as sensor noise, tracking deviations, dynamic environmental changes, and hardware experiments.
\bibliographystyle{IEEEtran} 
\bibliography{IEEEabrv,bibliography}

\begin{thebibliography}{10}
\providecommand{\url}[1]{#1}
\csname url@samestyle\endcsname
\providecommand{\newblock}{\relax}
\providecommand{\bibinfo}[2]{#2}
\providecommand{\BIBentrySTDinterwordspacing}{\spaceskip=0pt\relax}
\providecommand{\BIBentryALTinterwordstretchfactor}{4}
\providecommand{\BIBentryALTinterwordspacing}{\spaceskip=\fontdimen2\font plus
\BIBentryALTinterwordstretchfactor\fontdimen3\font minus
  \fontdimen4\font\relax}
\providecommand{\BIBforeignlanguage}[2]{{%
\expandafter\ifx\csname l@#1\endcsname\relax
\typeout{** WARNING: IEEEtran.bst: No hyphenation pattern has been}%
\typeout{** loaded for the language `#1'. Using the pattern for}%
\typeout{** the default language instead.}%
\else
\language=\csname l@#1\endcsname
\fi
#2}}
\providecommand{\BIBdecl}{\relax}
\BIBdecl

\bibitem{gu2018multiple}
J.~Gu, T.~Su, Q.~Wang, X.~Du, and M.~Guizani, ``Multiple moving targets
  surveillance based on a cooperative network for multi-uav,'' \emph{IEEE
  Communications Magazine}, vol.~56, no.~4, pp. 82--89, 2018.

\bibitem{peng2022obstacle}
P.~Peng, W.~Dong, G.~Chen, and X.~Zhu, ``Obstacle avoidance of resilient uav
  swarm formation with active sensing system in the dense environment,'' in
  \emph{2022 IEEE/RSJ International Conference on Intelligent Robots and
  Systems (IROS)}.\hskip 1em plus 0.5em minus 0.4em\relax IEEE, 2022, pp.
  10\,529--10\,535.

\bibitem{gao2022meeting}
Y.~Gao, Y.~Wang, X.~Zhong, T.~Yang, M.~Wang, Z.~Xu, Y.~Wang, Y.~Lin, C.~Xu, and
  F.~Gao, ``Meeting-merging-mission: A multi-robot coordinate framework for
  large-scale communication-limited exploration,'' in \emph{2022 IEEE/RSJ
  International Conference on Intelligent Robots and Systems (IROS)}.\hskip 1em
  plus 0.5em minus 0.4em\relax IEEE, 2022, pp. 13\,700--13\,707.

\bibitem{alcantara2021optimal}
A.~Alc{\'a}ntara, J.~Capit{\'a}n, R.~Cunha, and A.~Ollero, ``Optimal trajectory
  planning for cinematography with multiple unmanned aerial vehicles,''
  \emph{Robotics and Autonomous Systems}, vol. 140, p. 103778, 2021.

\bibitem{kim2016realization}
H.-J. Kim and H.-S. Ahn, ``Realization of swarm formation flying and optimal
  trajectory generation for multi-drone performance show,'' in \emph{2016
  IEEE/SICE International Symposium on System Integration (SII)}.\hskip 1em
  plus 0.5em minus 0.4em\relax IEEE, 2016, pp. 850--855.

\bibitem{zhao2021multi}
C.~Zhao, J.~Liu, M.~Sheng, W.~Teng, Y.~Zheng, and J.~Li, ``Multi-uav trajectory
  planning for energy-efficient content coverage: A decentralized
  learning-based approach,'' \emph{IEEE Journal on Selected Areas in
  Communications}, vol.~39, no.~10, pp. 3193--3207, 2021.

\bibitem{chen2021decentralized}
Y.~Chen, U.~Rosolia, and A.~D. Ames, ``Decentralized task and path planning for
  multi-robot systems,'' \emph{IEEE Robotics and Automation Letters}, vol.~6,
  no.~3, pp. 4337--4344, 2021.

\bibitem{I-Magnus2017ControlMRSsurvey}
J.~Cort{\'e}s and M.~Egerstedt, ``Coordinated control of multi-robot systems: A
  survey,'' \emph{SICE Journal of Control, Measurement, and System
  Integration}, vol.~10, no.~6, pp. 495--503, 2017.

\bibitem{cortes2017coordinated}
------, ``Coordinated control of multi-robot systems: A survey,'' \emph{SICE
  Journal of Control, Measurement, and System Integration}, vol.~10, no.~6, pp.
  495--503, 2017.

\bibitem{loquercio2021learning}
A.~Loquercio, E.~Kaufmann, R.~Ranftl, M.~M{\"u}ller, V.~Koltun, and
  D.~Scaramuzza, ``Learning high-speed flight in the wild,'' \emph{Science
  Robotics}, vol.~6, no.~59, p. eabg5810, 2021.

\bibitem{palossi201964}
D.~Palossi, A.~Loquercio, F.~Conti, E.~Flamand, D.~Scaramuzza, and L.~Benini,
  ``A 64-mw dnn-based visual navigation engine for autonomous nano-drones,''
  \emph{IEEE Internet of Things Journal}, vol.~6, no.~5, pp. 8357--8371, 2019.

\bibitem{miera2023lidar}
P.~Miera, H.~Szolc, and T.~Kryjak, ``Lidar-based drone navigation with
  reinforcement learning,'' \emph{arXiv preprint arXiv:2307.14313}, 2023.

\bibitem{el2023reinforcement}
S.~El-Ferik, M.~Maaruf, F.~Al-Sunni, A.~A. Saif, and M.~M. Al~Dhaifallah,
  ``Reinforcement learning-based control strategy for multi-agent systems
  subjected to actuator cyberattacks during affine formation maneuvers,''
  \emph{IEEE Access}, 2023.

\bibitem{liu2019reinforcement}
X.~Liu, Y.~Liu, and Y.~Chen, ``Reinforcement learning in multiple-uav networks:
  Deployment and movement design,'' \emph{IEEE Transactions on Vehicular
  Technology}, vol.~68, no.~8, pp. 8036--8049, 2019.

\bibitem{batra2022decentralized}
S.~Batra, Z.~Huang, A.~Petrenko, T.~Kumar, A.~Molchanov, and G.~S. Sukhatme,
  ``Decentralized control of quadrotor swarms with end-to-end deep
  reinforcement learning,'' in \emph{Conference on Robot Learning}.\hskip 1em
  plus 0.5em minus 0.4em\relax PMLR, 2022, pp. 576--586.

\bibitem{li2021message}
Q.~Li, W.~Lin, Z.~Liu, and A.~Prorok, ``Message-aware graph attention networks
  for large-scale multi-robot path planning,'' \emph{IEEE Robotics and
  Automation Letters}, vol.~6, no.~3, pp. 5533--5540, 2021.

\bibitem{li2020graph}
Q.~Li, F.~Gama, A.~Ribeiro, and A.~Prorok, ``Graph neural networks for
  decentralized multi-robot path planning,'' in \emph{2020 IEEE/RSJ
  International Conference on Intelligent Robots and Systems (IROS)}.\hskip 1em
  plus 0.5em minus 0.4em\relax IEEE, 2020, pp. 11\,785--11\,792.

\bibitem{blumenkamp2022framework}
J.~Blumenkamp, S.~Morad, J.~Gielis, Q.~Li, and A.~Prorok, ``A framework for
  real-world multi-robot systems running decentralized gnn-based policies,'' in
  \emph{2022 International Conference on Robotics and Automation (ICRA)}.\hskip
  1em plus 0.5em minus 0.4em\relax IEEE, 2022, pp. 8772--8778.

\bibitem{riviere2020glas}
B.~Riviere, W.~H{\"o}nig, Y.~Yue, and S.-J. Chung, ``Glas: Global-to-local safe
  autonomy synthesis for multi-robot motion planning with end-to-end
  learning,'' \emph{IEEE Robotics and Automation Letters}, vol.~5, no.~3, pp.
  4249--4256, 2020.

\bibitem{zhang2023neural}
S.~Zhang, K.~Garg, and C.~Fan, ``Neural graph control barrier functions guided
  distributed collision-avoidance multi-agent control,'' in \emph{Conference on
  Robot Learning}.\hskip 1em plus 0.5em minus 0.4em\relax PMLR, 2023, pp.
  2373--2392.

\bibitem{sabetghadam2022distributed}
B.~Sabetghadam, R.~Cunha, and A.~Pascoal, ``A distributed algorithm for
  real-time multi-drone collision-free trajectory replanning,'' \emph{Sensors},
  vol.~22, no.~5, p. 1855, 2022.

\bibitem{luis2020online}
C.~E. Luis, M.~Vukosavljev, and A.~P. Schoellig, ``Online trajectory generation
  with distributed model predictive control for multi-robot motion planning,''
  \emph{IEEE Robotics and Automation Letters}, vol.~5, no.~2, pp. 604--611,
  2020.

\bibitem{tordesillas2021mader}
J.~Tordesillas and J.~P. How, ``Mader: Trajectory planner in multiagent and
  dynamic environments,'' \emph{IEEE Transactions on Robotics}, vol.~38, no.~1,
  pp. 463--476, 2021.

\bibitem{kondo2023robust}
K.~Kondo, R.~Figueroa, J.~Rached, J.~Tordesillas, P.~C. Lusk, and J.~P. How,
  ``Robust mader: Decentralized multiagent trajectory planner robust to
  communication delay in dynamic environments,'' \emph{IEEE Robotics and
  Automation Letters}, 2023.

\bibitem{wang2022robust}
Z.~Wang, C.~Xu, and F.~Gao, ``Robust trajectory planning for spatial-temporal
  multi-drone coordination in large scenes,'' in \emph{2022 IEEE/RSJ
  International Conference on Intelligent Robots and Systems (IROS)}.\hskip 1em
  plus 0.5em minus 0.4em\relax IEEE, 2022, pp. 12\,182--12\,188.

\bibitem{park2022online}
J.~Park, D.~Kim, G.~C. Kim, D.~Oh, and H.~J. Kim, ``Online distributed
  trajectory planning for quadrotor swarm with feasibility guarantee using
  linear safe corridor,'' \emph{IEEE Robotics and Automation Letters}, vol.~7,
  no.~2, pp. 4869--4876, 2022.

\bibitem{bojarski2016visualbackprop}
M.~Bojarski, A.~Choromanska, K.~Choromanski, B.~Firner, L.~J. Ackel, U.~Muller,
  P.~Yeres, and K.~Zieba, ``Visualbackprop: Efficient visualization of cnns for
  autonomous driving,'' \emph{2018 IEEE International Conference on Robotics
  and Automation (ICRA)}, pp. 4701--4708, 2018.

\bibitem{tordesillas2022minvo}
J.~Tordesillas and J.~P. How, ``Minvo basis: Finding simplexes with minimum
  volume enclosing polynomial curves,'' \emph{Computer-Aided Design}, vol. 151,
  p. 103341, 2022.

\bibitem{chen2020learning}
D.~Chen, B.~Zhou, V.~Koltun, and P.~Kr{\"a}henb{\"u}hl, ``Learning by
  cheating,'' in \emph{Conference on Robot Learning}.\hskip 1em plus 0.5em
  minus 0.4em\relax PMLR, 2020, pp. 66--75.

\bibitem{charles2017pointnet}
R.~Q. Charles, H.~Su, M.~Kaichun, and L.~J. Guibas, ``Pointnet: Deep learning
  on point sets for 3d classification and segmentation,'' in \emph{2017 IEEE
  Conference on Computer Vision and Pattern Recognition (CVPR)}.\hskip 1em plus
  0.5em minus 0.4em\relax IEEE Computer Society, 2017, pp. 77--85.

\bibitem{tsitsulin2023graph}
A.~Tsitsulin, J.~Palowitch, B.~Perozzi, and E.~M{\"u}ller, ``Graph clustering
  with graph neural networks,'' \emph{Journal of Machine Learning Research},
  vol.~24, no. 127, pp. 1--21, 2023.

\bibitem{gama2022synthesizing}
F.~Gama, Q.~Li, E.~Tolstaya, A.~Prorok, and A.~Ribeiro, ``Synthesizing
  decentralized controllers with graph neural networks and imitation
  learning,'' \emph{IEEE Transactions on Signal Processing}, vol.~70, pp.
  1932--1946, 2022.

\bibitem{jiang2023c2spoint}
Z.~Jiang, L.~Ding, G.~K. Tam, C.~Song, F.~W. Li, and B.~Yang, ``C2spoint: A
  classification-to-saliency network for point cloud saliency detection,''
  \emph{Computers \& Graphics}, vol. 115, pp. 274--284, 2023.

\bibitem{zheng2019pointcloud}
T.~Zheng, C.~Chen, J.~Yuan, B.~Li, and K.~Ren, ``Pointcloud saliency maps,'' in
  \emph{Proceedings of the IEEE/CVF International Conference on Computer
  Vision}, 2019, pp. 1598--1606.

\bibitem{kingma2014adam}
D.~Kingma and J.~Ba, ``Adam: A method for stochastic optimization,'' in
  \emph{International Conference on Learning Representations (ICLR)}, San
  Diega, CA, USA, 2015.

\bibitem{amos2017optnet}
B.~Amos and J.~Z. Kolter, ``Optnet: Differentiable optimization as a layer in
  neural networks,'' in \emph{International Conference on Machine
  Learning}.\hskip 1em plus 0.5em minus 0.4em\relax PMLR, 2017, pp. 136--145.

\end{thebibliography}

\end{document}